\documentclass[12pt]{article}
\usepackage{amsmath,epsf,amssymb,latexsym,cite}
\newtheorem{prop}{Proposition}

\newif\iffigs\figstrue

\DeclareFontFamily{U}{rsf}{}
\DeclareFontShape{U}{rsf}{m}{n}{
  <5> <6> rsfs5 <7> <8> <9> rsfs7 <10-> rsfs10}{}
\DeclareMathAlphabet\Scr{U}{rsf}{m}{n}

\def\pplogo{\vbox{\kern-\headheight\kern -29pt
\halign{##&##\hfil\cr&{
\ppnumber}\cr\rule{0pt}{2.5ex}&\ppdate\cr}
}}
\makeatletter
\def\ps@firstpage{\ps@empty \def\@oddhead{\hss\pplogo}%
  \let\@evenhead\@oddhead 
}
\def\maketitle{\par
 \begingroup
 \def\thefootnote{\fnsymbol{footnote}}
 \def\@makefnmark{\hbox{$^{\@thefnmark}$\hss}}
 \if@twocolumn
 \twocolumn[\@maketitle]
 \else \newpage
 \global\@topnum\z@ \@maketitle \fi\thispagestyle{firstpage}\@thanks
 \endgroup
 \setcounter{footnote}{0}
 \let\maketitle\relax
 \let\@maketitle\relax
 \gdef\@thanks{}\gdef\@author{}\gdef\@title{}\let\thanks\relax}
\makeatother

\def\P{{\mathbb P}}

\def\R{{\mathbb R}}
\def\Z{{\mathbb Z}}

\def\Img{\operatorname{Im}}

\def\SO{\operatorname{SO}}
\def\Sl{\operatorname{SL}}
\def\GO{\operatorname{O{}}}
\def\SU{\operatorname{SU}}
\def\GU{\operatorname{U{}}}
\def\Sp{\operatorname{Sp}}

\def\CY{Calabi--Yau}

\def\cM{{\Scr M}}

\def\cT{{\Scr T}}

\def\RoR{$R\leftrightarrow1/R$}

\begin{document}
\setcounter{page}0
\def\ppnumber{\vbox{\baselineskip14pt\hbox{DUKE-CGTP-99-04}
\hbox{hep-th/9905036}}}
\def\ppdate{May 1999} \date{}

\title{\LARGE T-Duality Can Fail\\[10mm]}
\author{
Paul S. Aspinwall and M. Ronen Plesser\\[10mm]
\normalsize Center for Geometry and Theoretical Physics, \\
\normalsize Box 90318, \\
\normalsize Duke University, \\
\normalsize Durham, NC 27708-0318\\[10mm]
}

{\hfuzz=10cm\maketitle}

\def\Large{\large}
\def\LARGE{\large\bf}


\begin{abstract}
We show that T-duality can be broken by nonperturbative effects in
string coupling. The T-duality in question is that of the 2-torus when
the heterotic string is compactified on K3$\times T^2$.  This case is
compared carefully to a situation where T-duality appears to work.
A holonomy argument is presented to show that T-dualities (and general
U-dualities) should only be expected for large amounts of supersymmetry.
This breaking of \RoR\ symmetry raises some interesting questions in
string theory which we discuss. Finally we discuss how the classical
modular group of a 2-torus appears to be broken too.

\end{abstract}

\vfil\break


\section{Introduction}    \label{s:int}

\RoR\ symmetry, the statement that a string theory compactified on a
circle of radius $R$ is isomorphic to a string compactified on a
circle of radius $1/R$, is the most basic instance of T-duality.  It
is not only a prime example of inherently ``stringy'' physics but also
an extremely useful tool in studying the theory.  It has been
used to argue for the existence of ``minimal distances'' in string
theory and, applied in more complicated situations, to derive 
various properties of the theory.  In these applications it is often
necessary to assume that T-duality is an exact symmetry of string
theory. The justification of \RoR\ symmetry
comes from conformal field theory. Probably the neatest argument is
that if one looks at a conformal field theory corresponding to a
string on a circle of radius $R$, then at $R=1$ (in suitable units)
one gets an extra $\SU(2)$ symmetry which may be used to identify the
marginal operator which decreases $R$ with the marginal operator which
increases $R$ (see, e.g., \cite{Gins:lect}).  In compactification on
$T^2$ the T-duality is extended to an $\Sl(2,\Z)$ symmetry acting on
the volume and background $B$-field.

The problem with the argument is that it ignores nonperturbative
effects in the string coupling constant. In this paper we will show
how nonperturbative effects can modify the conclusion when the string
coupling is nonzero.  Our general conclusion is that T-duality-like
arguments work generally only when there is a lot of supersymmetry.
Here we will look at the heterotic string compactified on a product of
a K3 surface and a 2-torus leading to an $N=2$ theory in four
dimensions.  Heterotic strings compactified on $T^2$ exhibit T-duality
at any value of the string coupling, as we will show.  At the level of
conformal field theory the correlation functions factorize into
products of correlators associated to one or the other factor of the
compactification space, so the presence of K3 is irrelevant to the
issue of T-duality for $T^2$.  We will show that this is not true
nonperturbatively.  The 2-torus in this context appears to be the most
supersymmetric case for a failure of T-duality and so will be the
easiest to analyze.

In many ways the essential facts of this paper have been known to many
people for a few years now. That is, something peculiar happens to
T-duality when a heterotic string is compactified on K3$\times T^2$.
This was studied in particular in \cite{CDFvP:,AFGNT:mono,HM:count} for example.
Our purpose here is to emphasize that this implies a basic failure of
T-duality in a general setting --- a fact which does not appear to
have been widely appreciated.

We should immediately clarify precisely what we mean when we say
that T-duality is ``broken''. What we do {\em not\/} mean is that we
consider the string theory on a circle of radius $R$ and then the same
string theory on a circle of radius $1/R$ and discover that they are
different. The situation is more awkward than this because it is
difficult to say unambiguously how to measure the radius of a circle in
this context.  

What we can describe unambiguously is the moduli space of heterotic
string theories compactified on a given spacetime.  We can then follow
the philosophy developed in studying moduli spaces of \CY\ threefolds
\cite{CDGP:,AGM:sd,me:min-d}.  We attempt to assign a ``geometrical''
interpretation in terms of parameters such as the sizes of circles to
points in the moduli space. One begins with some large radius (and/or weak
coupling) region where one understands the system classically and
then one integrates along paths in the moduli space assigning
parameters to all the points in the moduli space. The problem with
this method is that monodromies within the moduli space force branch
cuts to be made. In general one finds that this process
requires choices, so that the parameters are not uniquely
determined by the moduli. 
One may try to remove these monodromies by forming a Teichm\"uller
space which will cover this moduli space.
If all goes well then the new copies of the
fundamental region will tessellate ``naturally'' to form a cover giving some
parameter space. An example where this fails was first recognized for
measuring the volume of the quintic threefold in \cite{CDGP:} (see
figure 5.2 of that paper in particular).

It is important to note that one may always form the Teichm\"uller
space and obtain an associated (usually very large) modular
group. Indeed this was the 
approach taken in \cite{CDFvP:} for example and one may also obtain
useful information about the theory if this path is taken (see
\cite{HM:count} for example). The point we wish to emphasize here
however is that this Teichm\"uller space is often physically
meaningless in the usual sense. 

What conditions need to hold in order for a physical interpretation
to survive?  In this paper we propose the following.  The parameter
space must contain a limiting ``semiclassical'' open region corresponding to
weakly coupled strings and large radii.  This region should be unique
up to the action of duality transformations which can be determined
from the semiclassical approximation to the system.  In the case at
hand this would mean that the large-radius, weak-coupling limit of a
$T^2$ compactification must be unique up to $B\to B+1$.  This
requirement is quite restrictive as we shall see. 

We believe this is a very reasonable definition for T-duality to
respect. One could not seriously propose that there is a symmetry in the
universe which identifies a circle of radius 1cm with a circle of
radius 2cm --- we may use a ruler to measure the difference! An
important aspect of T-dualities is that large distances are unique. It
is precisely this uniqueness which is lost if we allow an arbitrary
cover of the moduli space to act as our Teichm\"uller space.

We will endeavour to show that the moduli space for the 2-torus in the
context above does not admit a covering with this property and so
there is no T-duality group. Note that \RoR\ symmetry is not literally 
``broken''. Instead what is true is that for $R$ of order one or
smaller, there is no natural way to associate a ``size'' $R$ to a
given point in moduli space.  In particular, the statement of
T-duality thus loses its content.

The nonperturbative corrections responsible for modifying the
structure of the moduli space and rendering it inconsistent with
T-duality are presumably represented by fivebrane instantons wrapping
the K3 as well as one of the cycles of the torus.  We do not know at
present how to compute these corrections explicitly.  Instead, we use
a dual type IIA description of the model.  In this description, the
heterotic dilaton is mapped to a K\" ahler modulus and the corrections
in question are generated by world-sheet instantons whose effect is
readily computed by mirror symmetry.  

In the context of this dual model the issue of T-duality breaking
becomes that of {\em fibre-wise
duality}. The idea is that one has a \CY\ space, $X$, which is an
$F$-fibration for some fibre $F$ which is also a \CY\ space. If some
duality statement is true for each $F$ can it be extended to a duality
of $X$? In section \ref{s:K3} we give an example where it can and in
section \ref{s:snap} we give an example where it cannot. By
heterotic/type II string duality the latter case implies a failure of
T-duality. The essential difference between sections \ref{s:K3} and
\ref{s:snap} is that in former we show that we are required to fit a
{\em finite\/} number of fundamental domains of the moduli space into
a part of the Teichm\"uller space with {\em infinite\/} area while in
the later case we would be required to fit an
{\em infinite\/} number of fundamental domains of the moduli space into
a part of the Teichm\"uller space with {\em finite\/} area.

In section \ref{s:conc} we will discuss the consequences of broken
dualities. In particular we will set a general description of how
dualities are natural only when there is a large amount of
supersymmetry.  We propose that the clearest interpretation of the
results in this paper is that {\em whether a circle respects T-duality
in string theory depends on the context of the circle.}  We will also
note the possibly alarming conclusion that the {\em classical\/}
$\Sl(2,\Z)$ modular group of the moduli space of complex structures on
a torus is also probably broken (in the same sense).


\section{Unbroken T-Duality}  \label{s:K3}

\subsection{A Two Parameter K3 Surface}  \label{ss:K2}

In this section we will deal with a type IIA string compactified on a
K3 surface which is an elliptic fibration with section. We wish to see
that the $\Sl(2,\Z)$ symmetry of the elliptic fibre can be seen in the
moduli space of the K3 surface. 

An F-theory argument shows that this is equivalent to saying that the
$\Sl(2,\Z)$ T-duality acting on the moduli space of complexified
K\"ahler forms is unaffected by the value of the complex structure
modulus (or, equally, the mirror of this statement). Note that the
heterotic dilaton in eight dimensions lives in a separate $\R$ factor
of the moduli space and will certainly not affect any T-duality statements.

We know from Narain
\cite{N:torus} that the moduli space required for a heterotic string
on $T^2$ (with the Wilson lines switched off) is
\begin{equation}
  \cM_0 = \GO(\Gamma_{2,2})\backslash\GO(2,2)/(\GO(2)\times\GO(2)),
	\label{eq:M0}
\end{equation}
where $\Gamma_{2,2}=U\oplus U$ and $U$ is the usual hyperbolic lattice
of signature $(-1,1)$. This space can be thought of as the Grassmannian
of space-like 2-planes in $\R^{2,2}$ where we care only about the
orientation of the 2-planes relative to the lattice $\Gamma_{2,2}$.

One may explicitly map $\Sl(2,\R)\times\Sl(2,\R)$ to $\GO(2,2)$ by
writing a vector $(x_1,x_2,x_3,x_4)$ in $\R^{2,2}$ as
\begin{equation}
  M = \begin{pmatrix} x_1&x_3\\ x_4&x_2 \end{pmatrix},
\end{equation}
and then let $(A,B)\in\Sl(2,\R)\times\Sl(2,\R)$ act on this vector as
\begin{equation}
  (A,B): M\to AMB^{-1}.
\end{equation}
(Note that $(-1,-1)$ maps to the identity.)

This allows us to rewrite $\cM_0$ in (\ref{eq:M0}) in terms of two
copies of the upper-half plane $\Sl(2,\R)/\GU(1)$. The modular group
$\GO(\Gamma_{2,2})$ then generates two copies of $\Sl(2,\Z)$ --- each
acting on its upper-half plane, a ``mirror'' $\Z_2$ which exchanges
the upper-half planes, and a ``conjugation'' $\Z_2$ which acts as
(minus) complex conjugation on both upper-half planes simultaneously.
One can then interpret one upper-half plane as representing the
complex structure, $\tau$, of the 2-torus and the other upper-half
plane as representing the $B$-field and K\"ahler form, $\sigma= B+iJ$,
of the torus \cite{DVV:torus}. Equivalently we may describe this
moduli space as two copies of the $j$-line divided by exchange and
complex conjugation. We denote coordinates on the two $j$-lines by
$j_1$ and $j_2$.

Via the usual F-theory argument \cite{Vafa:F} the heterotic string on
$T^2$ is dual to F-theory (or the type IIA string in some limit) on a
K3 surface, $S$.\footnote{There is a very awkward $\Z_2$
identification problem which we will ignore for the sake of
exposition. The complex conjugation symmetry of $\cM_0$ must be
treated carefully in the type IIA picture. We will ignore this
subtlety for the sake of exposition.}
We will map the moduli space
explicitly for the case of an algebraic K3 surface written as a
hypersurface in a toric variety. Let us consider the surface $\tilde
S$ given by 
\begin{equation}
  x_0^2 + x_1^3 + x_2^{12} + x_3^{12} + \psi x_0x_1x_2x_3
	+\phi x_2^6x_3^6,     \label{eq:K3a}
\end{equation}
in $\P_{\{6,4,1,1\}}^3/\Z_6$, where we divide by the $\Z_6$ action
generated by $(x_0,x_1,x_2,x_3)\mapsto(x_0,x_1,\allowbreak e^{\pi
i/6}x_2,\allowbreak e^{-\pi i/6}x_3)$.\footnote{Although this action
looks like it 
generates $\Z_{12}$, don't forget that the weighted projective
identification means that $(x_0,x_1,x_2,x_3)\cong(x_0, x_1,-x_2, -x_3)$.}
This K3 surface is mirror via the usual arguments \cite{GP:orb,AM:K3p}
to an elliptic K3 surface with a section. 
That is, $S$ and $\tilde S$ are a mirror pair in the sense of
algebraic K3 surfaces \cite{AM:K3p,Dol:K3m,me:lK3}.
We should now be able
to map out the desired moduli space, $\cM_0$ by varying $\psi$ and
$\phi$ in (\ref{eq:K3a}), i.e., by varying the complex structure of
$\tilde S$.

We want to explicitly map the coordinates $(\psi,\phi)$ to the two
copies of the $j$-line we saw above. 
This has already been done in \cite{LY:K3f,KLM:K3f} (see also
\cite{CCLM:8E}). Let us review this construction so that we may
compare some details to that of the \CY\ threefold in the next section.
The structure is most easily seen by finding the ``interesting''
points in the moduli space. As is well-known, at certain points in the
moduli space we obtain enhanced gauge symmetries. In the language of the
Grassmannian (\ref{eq:M0}) this occurs when lattice elements of length
squared $-2$ are orthogonal to the space-like 2-plane. Let us suppose
that we force the 2-plane to be orthogonal to such a lattice
element. The result is that the 2-plane now varies in the space
$\Gamma_{2,1}\otimes_\Z\R$, where $\Gamma_{2,1}=U\oplus L(2)$ and
$L(2)$ is a one-dimensional lattice whose generator has length
squared $2$. That is, the moduli space within $\cM_0$ where the
theory has a gauge symmetry of at least $\SU(2)$ is given by
\begin{equation}
  \cM_{\SU(2)} = \GO(\Gamma_{2,1})\backslash\GO(2,1)/(\GO(2)\times\GO(1)).
	\label{eq:Msu2}
\end{equation}
In the same way that $\GO(2,2)$ is mapped to
$\Sl(2,\R)\times\Sl(2,\R)$, we may map $\GO(2,1)$ to $\Sl(2,\R)$ by 
writing
\begin{equation}
  M = \begin{pmatrix} x_1&x_2\\ x_3&-x_1 \end{pmatrix},
\end{equation}
and letting $A\in\Sl(2,\R)$ act as
\begin{equation}
  A: M\to AMA^{-1}.
\end{equation}
The reader may also check that (up to a $\Z_2$ corresponding to
complex conjugation) $\GO(\Gamma_{2,1})\cong\Sl(2,\Z)$.

Thus we may embed $\cM_{\SU(2)}$ into $\cM_0$ by setting $A=B$. In
other words $j_1=j_2$ in terms of the two $j$-lines. Actually 
the fact that $\GO(\Gamma_{2,2})$ acts {\em transitively\/} on vectors
of length squared $-2$ shows that we have
an enhanced gauge symmetry (of at least $\SU(2)$) if {\em and only
if\/} $j_1=j_2$.

In terms of $\tilde S$ in the form (\ref{eq:K3a}) we have an enhanced gauge
symmetry when $\phi$ and $\psi$ are tuned to produce a canonical
singularity. This happens when the discriminant, $\Delta$, of the
equation vanishes. This is given by\footnote{Note that there are some
identifications to be made within the $(\psi,\phi)$ plane. This
discriminant does not really have four components.}
\begin{equation}
  \Delta = (\phi  - 2)(\phi  + 2)(432\phi  - \psi ^{6} + 
  864)(432\phi  - \psi ^{6} - 864).
\end{equation}
One may also argue that $\Delta=k(j_1-j_2)^2$, where $k$ is some
constant, as follows. As we said above we only get enhanced gauge
symmetry when $j_1=j_2$. One argues that the power is 2 by saying that
any odd power would violate the $j_1\leftrightarrow j_2$ symmetry and
any power higher than 2 would imply that the degeneration of $\tilde S$ when
$\Delta=0$ could never be suitably generic.

We also have further enhanced gauge symmetry. Namely there is a single
point $j_1=j_2=0$ where we have $\SU(3)$ and a
single point $j_1=j_2=1728$ where we have
$\SU(2)\times\SU(2)$. With a bit of algebra we may find the
corresponding values for $\phi$ and $\psi$.
All said, our equations are satisfied with $k=1$ by making $j_1$ and
$j_2$ the roots of
\begin{equation}
  j^2 - (\phi\psi^6-432\phi^2+1728)\,j + \psi^{12}=0.
	\label{eq:j1}
\end{equation}

If the reader is not fully convinced of this map between $(\psi,\phi)$
and the $j$-lines, one may check that the ``flatness'' condition is
satisfied for the ``special coordinates'' $\sigma$ and $\tau$. 
We will do this next. For a
nice account of what is meant by ``special coordinates'' we refer to
\cite{Frd:SK}.

The special coordinates are of particular interest to us because these
specify ``length'' are should parameterize any natural Teichm\"uller
space. Let us perform the usual analysis of converting ``algebraic
coordinates'' $(\psi,\phi)$ into special, or flat coordinates as 
in \cite{CDGP:,GZK:h,AGM:sd}. This is a rather technical process but
it is thoroughly treated in the literature. Our approach is closest to
\cite{me:min-d}.
Introducing variables
\begin{equation}
  z = \frac{\phi}{\psi^6},\quad\quad y = \frac1{\phi^2},
\end{equation}
we obtain the hypergeometric partial differential operators
\def\ldiff#1{#1\frac{\partial}{\partial#1}}
\begin{equation}
\begin{split}
  \Box_z &= \ldiff z\left(\ldiff z-2\ldiff y\right)-432\,z\left(\ldiff
  z+\frac16\right)\left(\ldiff z+\frac56\right)\\
  \Box_y &= \left(\ldiff y\right)^2 - y\left(\ldiff z-2\ldiff y\right)
   \left(\ldiff z-2\ldiff y-1\right)
\end{split}
	\label{eq:hs2}
\end{equation}
The monomial-divisor mirror map of \cite{AGM:mdmm} tells us that $S$
is an elliptic fibration with a section and that the section goes to
infinite size as $y\to0$ and the fibre goes to infinite size as
$z\to0$. There is a unique solution of
$\Box_z\Phi(z,y)=\Box_y\Phi(z,y)=0$ which is equal to 1 at the
origin. Call this $\Phi_0$. We also have solutions $\Phi_z$ and
$\Phi_y$ which behave asymptotically as $\log(z)$ and $\log(y)$ as we
approach the origin. The flat coordinates are then ratios of these
solutions in the usual way. Let us put
\begin{equation}
q_z=\exp\left(\frac{\Phi_z}{\Phi_0}\right),\qquad
q_y=\exp\left(\frac{\Phi_y}{\Phi_0}\right).
\end{equation}
(Just as $q$ is used
conventionally to denote $\exp(2\pi i\tau)$.) We then have a power
series solution from the hypergeometric system:
\begin{equation}
\begin{split}
  q_z &= z-zy+312z^2-zy^2-192z^2y+107604z^3+\ldots\\
  q_y &= y + 120zy+2y^2+41580yz^2+5y^3+\ldots
\end{split}
\end{equation}
Explicit computation then shows order by order that this is consistent
with (\ref{eq:j1}) with 
\begin{equation}
\begin{split}
  j_1 &= j(q_z)\\
  j_2 &= j(q_zq_y),
\end{split}
\end{equation}
where $j(q)$ is the usual $j$-function expansion given by $q^{-1}+744
+196884q+O(q^2)$. This verifies that (\ref{eq:j1}) is correct.
See \cite{LY:K3f} for an explicit proof of this.

\subsection{Fibre-wise T-Duality}  \label{ss:fibre}

Now we wish to analyze the same two parameter model from the point of
view of the elliptic fibration of the K3 surface $S$. 
We want to know if $\Sl(2,\Z)$ modular group of the elliptic fibre can
be seen in the moduli space of $S$. First let us take the area of
section to be huge, $y\to0$.  
In this limit,
the moduli properties of $S$ are basically that of its fibre. Putting
$y=0$ we obtain
\begin{equation}
  j_1=j(q_z)=\frac1{z(1-432z)},
		\label{eq:j1z}
\end{equation}
and $j_2=\infty$. It is very important to note that varying $z$ gives
us a {\em double\/} cover of the $j_1$-line. In particular, if we plot
the shape of fundamental region in terms of flat coordinates (as
discussed in \cite{CDGP:,AGM:sd,me:min-d}) we obtain the shaded area
of figure \ref{fig:G2}. This region is $H/\Gamma_0(2)$, where $H$ is
the upper-half plane. This means that there is a non-toric $\Z_2$ symmetry
acting on $\tilde S$ which will identify the two fundamental regions
of $H/\Sl(2,\Z)$ in figure \ref{fig:G2}.

\iffigs
\begin{figure}
  \centerline{\epsfxsize=7cm\epsfbox{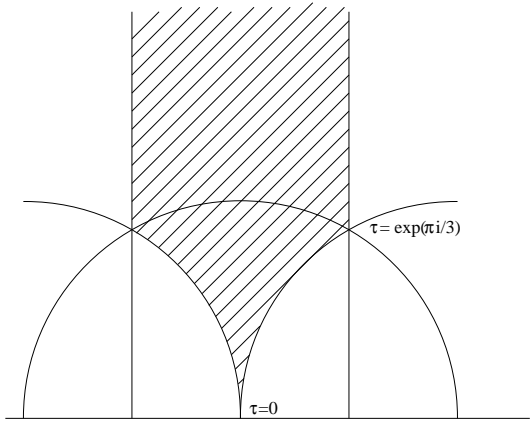}}
  \caption{The domain of mapping into flat coordinates when $y=0$.}
  \label{fig:G2}
\end{figure}
\fi
 
In terms of the mirror $\tilde S$, taking $y\to0$ takes
$\phi\to\infty$. $\tilde S$ is an elliptic K3 surface with a section
and in this limit the complex structure of the fibre becomes
constant. The remaining variation of complex structure of $\tilde S$
is therefore the variation of the complex structure of this constant
fibre. A similar story was was discussed in \cite{VW:pairs}.

Finding the moduli space of an elliptic curve via toric varieties is
always going to be plagued by finding multiple covers of the $j$-line
because of non-toric symmetries. Most simple families
of elliptic curves that one tries to write down tend to have fixed
rational points. See \cite{LLW:ell} for a discussion of
such problems. This ``level structure'' forces the na\"\i ve moduli
space to be a multiple cover of the $j$-line. This helps us out in the
toric case as we now explain.

In the general approach of Candelas et al.~\cite{CDGP:} in analyzing a
one parameter case, the moduli space will always look
like a rational curve with 3 special points. 
To use the language of the quintic threefold,
one point will be ``large
complex structure'' another will be an orbifold-like point and a
third, which marks the division between the first two phases, is
a ``conifold''-like point. The problem is that an elliptic curve
doesn't have a conifold-like point --- any singular complex structure
must be the large complex structure. Thus one must have a non-toric
symmetry identifying the putative conifold point with the large complex
structure point. In our example this must identify the ``conifold''
$\tau=0$ with $\tau=i\infty$, i.e., it is the $\tau\to-1/\tau$
symmetry. It is this non-toric discrete symmetry which produces the
multiple cover of the $j$-line above.

Now the main question we wish to ask is what happens to figure
\ref{fig:G2} if we allow $y$ to be nonzero? From (\ref{eq:j1z}) the
point given by $\tau=0$ in figure \ref{fig:G2} corresponds to
$z=1/432$. This is on the discriminant as it corresponds to
$j_1=j_2=\infty$. For general $y$, the relevant part of the
discriminant becomes 
\begin{equation}
(432\,z-1)^2-864^2\,yz^2=0.   \label{eq:dc1}
\end{equation}
Thus, as $y$ is switched
on, the point $z=1/432$ splits into two nearby points (up to extra
identifications). How can we understand this in terms of flat
coordinates?

The question of what a slice of constant $y$ in the moduli space
corresponds to in terms of the elliptic curve is awkward. Recall that
setting $y=0$ put $j_2=\infty$ and so $\sigma=i\infty$. Suppose
instead we fix $\sigma$ to be $iC$ for some large but finite $C$. What
does the fundamental region for $\tau$ then look like? Because of the
mirror map $\sigma\leftrightarrow\tau$, we are free to fix $\Img(\tau)
\leq\Img(\sigma)$. This ``chops off'' the top of the fundamental
region. This will turn figure \ref{fig:G2} into figure \ref{fig:G2a}.

\iffigs
\begin{figure}
  \centerline{\epsfxsize=5cm\epsfbox{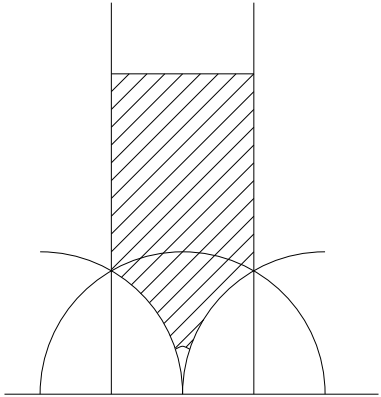}}
  \caption{The domain when $\sigma=iC$.}
  \label{fig:G2a}
\end{figure}
\fi

Now the domain for $y$ constant, as opposed to $j_2$ constant, won't
quite look like figure 
\ref{fig:G2a} but the coarse structure should be about right. Thus we
see how the point at the end of the cusp at $\tau=0$ in figure
\ref{fig:G2} gets split and moves up a little in figure
\ref{fig:G2a}. (We should also point out that these supposed two points
are actually identified by the $\Sl(2,\Z)$ action in figure
\ref{fig:G2a}.)
When $y$ is nonzero, a solution of $z$ satisfying (\ref{eq:dc1}) gives
a $\Z_2$-orbifold point on $S$. Thus this corresponds to an $\SU(2)$
gauge symmetry (in the eight-dimensional theory in question) 
and this point in the moduli space is a $\Z_2$ orbifold point.

The important point is that {\em nothing catastrophic happens to the
$\Sl(2,\Z)$ action as we switch on $y$}. 
We just pick up an extra $\Z_2$ into the modular group.
We know that our moduli space
is of the form of a global quotient (\ref{eq:M0}) and so no matter
what kind of complex slice of the moduli space we take, we will always
see some T-duality group acting nicely on some slice of the
Teichm\"uller space. In the heterotic string point of view, the T-duality
group of the torus persists.


\section{Broken T-Duality}    \label{s:snap}

We now wish to discuss along parallel lines the case of a heterotic
string compactified on K3$\times T^2$.  To this end, we need to 
replace the K3 surface $S$ with a \CY\ threefold
$X$. The space $X$ is now a K3-fibration. The fibre has a T-duality
which can be shown to be $\Sl(2,\Z)$ (which up to $\Z_2$ factors is
$\GO(\Gamma_{1,2})\subset\GO(\Gamma_{4,20})$ --- the full T-duality
group of a K3 
surface). This $\Sl(2,\Z)$ will not survive to the full \CY\ threefold
moduli space. This shows that the T-duality of the
heterotic string is destroyed at nonzero string coupling.

The toric data for $X$ will be very similar to that of $S$, and
$X$ will again have two deformations of the $B$-field and K\"ahler
form. Let $\tilde X$ be given by
\begin{equation}
  x_0^2 + x_1^6 + x_2^6 + x_3^{12} + x_4^{12} + \psi x_0x_1x_2x_3x_4
	+\phi x_3^6x_4^6,     \label{eq:Xa}
\end{equation}
in $\P_{\{6,2,2,1,1\}}^3/G$, where we divide by the
$G=\Z_6\times\Z_6\times\Z_2$ action
generated by
\begin{equation}
\begin{split}
(x_0,x_1,x_2,x_3,x_4)&\mapsto(x_0,x_1,x_2,e^{\pi i/6}x_3,e^{-\pi i/6}x_4)\\
(x_0,x_1,x_2,x_3,x_4)&\mapsto(x_0,x_1,e^{\pi i/3}x_2,x_3,e^{-\pi i/3}x_4)\\
(x_0,x_1,x_2,x_3,x_4)&\mapsto(x_0,e^{\pi i/3}x_1,x_2,x_3,e^{-\pi i/3}x_4).
\end{split}
\end{equation}
This \CY\ threefold and its mirror $X$ have been studied extensively
\cite{CDFKM:I}. In particular Kachru and Vafa \cite{KV:N=2}
conjectured that the type IIA string on $X$ is dual to the heterotic
string on $S_H\times E_H$ where $S_H$ is a K3 surface and $E_H$ is an
elliptic curve. The string theory on $E_H$ is frozen along
$\tau=\sigma$ by Higgsing the enhanced $\SU(2)$ symmetry we saw in the
last section. Thus the moduli space of the string on $E_H$ should be just a
single copy of the $j$-line. There is considerable evidence to support
this conjecture (for example \cite{KV:N=2,KKL:limit,DKLL:}).

Now as before we may introduce 
\begin{equation}
  z = \frac{\phi}{\psi^6},\quad\quad y = \frac1{\phi^2},
\end{equation}
and we obtain the hypergeometric partial differential operators
\def\ldiff#1{#1\frac{\partial}{\partial#1}}
\begin{equation}
\begin{split}
  \Box_z &= \left(\ldiff z\right)^2
	\left(\ldiff z-2\ldiff y\right)-1728\,z\left(\ldiff
  z+\frac16\right)\left(\ldiff z+\frac12\right)\left(\ldiff z+\frac56\right)\\
  \Box_y &= \left(\ldiff y\right)^2 - y\left(\ldiff z-2\ldiff y\right)
   \left(\ldiff z-2\ldiff y-1\right)
\end{split}
	\label{eq:hs3}
\end{equation}
Again the monomial-divisor mirror map of \cite{AGM:mdmm} tells us that $X$
is a K3 fibration with a section and that the section goes to
infinite size as $y\to0$ and the K3 fibre goes to infinite size as
$z\to0$. Note that this hypergeometric system is remarkably similar to the
elliptic fibration in the previous section (\ref{eq:hs2}).

We may find the flat coordinates as before, the exponentials of which we
will call $q_z$ and $q_y$ again. Now when we put $y=0$ one finds
\cite{CDFKM:I}
\begin{equation}
  j(q_z) = \frac1z.
\end{equation}
That is, the moduli space of the K3 fibre gives one cover of the
$j$-line. The fact that the $j$-line appears here can be argued
directly as follows. With a suitable change of coordinates, the
constant K3 fibre of (\ref{eq:Xa}) as $y\to0$ may be written as
\begin{equation}
  x_0^2 + x_1^6 +x_2^6+x_3^6 + \xi\,x_2^2x_3^2x_4^2,
\end{equation}
in $\P_{\{3,1,1,1\}}^3$ divided by $\Z_2\times\Z_6$. This is
manifestly a double cover of $\P^2$ branched over a sextic
curve. We may write $\P^2$ as a $\Z_2\times\Z_2$-cover of $\P^2$ by
mapping $[y_1,y_2,y_3]\to[x_1^2,x_2^2,x_3^2]$. This $\Z_2\times\Z_2$
is a subgroup of the $\Z_2\times\Z_6$ by which we are going to
orbifold. That is, the sextic curve is mapped to a cubic --- i.e., an
elliptic curve. Thus the moduli space of the K3 surface can be mapped
to the moduli space of an elliptic curve. A little work is involved in
showing that there is no level structure implicit on this elliptic
curve and so one indeed gets one copy of the $j$-line as desired.

The heterotic interpretation of the flat coordinates corresponding to
$q_z$ and $q_y$ are the single complex modulus of the torus and the
dilaton-axion respectively \cite{KV:N=2}. Indeed, the appearance of
the $j$-line above as $y\to0$ shows that, in this limit we do indeed
get the correct moduli space for the torus --- $H/\Sl(2,\Z)$. In this
way we see that the T-duality of the heterotic string on a torus at
zero coupling is reproduced in the type-IIA dual.

Now what happens when we allow $y$ to be nonzero? Although the
hypergeometric system (\ref{eq:hs3}) is similar at first sight to that
of the previous section (\ref{eq:hs2}), there is a drastic
difference. Let us consider the behaviour of the discriminant locus in
a constant $y$ slice as $y$ is near zero. As discussed above, a
component near $z=1/432$ gave us two points (before identifications
are made) close to where a {\em cusp\/} of the fundamental region was in the
$y=0$ limit. This cusp corresponded to $j=\infty$.

In this case in question here, the corresponding component of the
discriminant locus is given by
\begin{equation}
  (1-1728\,z)^2-3456^2\,z^2y=0.
\end{equation}
At $y=0$, the point $z=1/1728$ is not a cusp on the $j$-line but
rather the $\Z_2$-orbifold point at $j=1728$. In the language of the
K3 fibre of $X$, the K3 surface acquires a $\Z_2$-orbifold point here
(along with $B=0$) to give an enhanced $\SU(2)$ gauge symmetry. 

Now as we turn on $y$ we have two solutions for $z$. This is exactly
the situation that was analyzed thoroughly by Kachru et al
\cite{KV:N=2,KKL:limit} where it was shown that the region of
moduli space near $z=1/1728$ and $y=0$ maps to Seiberg-Witten theory
for an $\SU(2)$ theory \cite{SW:I}. The single point $y=0$, $z=1/1728$
is the classical limit corresponding to an $\SU(2)$ enhanced gauge
symmetry and the two points for constant $y>0$ are the points giving
the massless solitons. The important result we wish to borrow from
Seiberg-Witten theory is that in flat coordinates the monodromy around
either of these two points alone is {\em infinite\/}. The monodromy
within the curve $y=0$ around both points together is $\Z_2$ --- as we
see from the $y=0$ limit.\footnote{This is not to say that the
monodromy around the two points in the Seiberg--Witten plane is
$\Z_2$. See \cite{KKL:limit}.}

The situation here contrasted with that of the previous section can be
summed up as follows
\begin{itemize}
  \item For the moduli space of $S$ we have a point at $y=0$ around
  which the monodromy is infinite. As $y$ is switched on this splits
  into two points around which the monodromy is $\Z_2$.
  \item For the moduli space of $X$ we have a point at $y=0$ around
  which the monodromy is $\Z_2$. As $y$ is switched on this splits
  into two points around which the monodromy is infinite.
\end{itemize}
This can all be traced back to the fact that for $S$, $y=0$ gave a
double cover of the $j$-line while for $X$ it gives a single cover.

We now wish to claim that this behaviour for $X$ makes it completely
impossible to maintain any notion of T-duality when $y$ is switched
on. That is T-duality for the heterotic string is ``broken'' as soon
as the dilaton is switched on.

Suppose we let $y$ be small but nonzero. This should correspond to the
weakly-coupled heterotic string. The moduli space for the $T^2$ part
must look ``almost'' like the classical fundamental region for
$H/\Sl(2,\Z)$. By ``almost'' we mean that no points in the moduli
space for $T^2$ can have moved a large distance with respect to the special
K\"ahler metric relative to the $y=0$ slice. We should now ask whether
we can apply $\Sl(2,\Z)$ to this new moduli space to form a nice
tessellated cover of the upper half plane just as we could for $y=0$.
Note in particular the the angle of any corner of the
moduli space must correspond to the monodromy in the Levi-Civita
connection of the special K\"ahler metric around that point in order
that such a tessellation works.

\iffigs
\begin{figure}
  \centerline{\epsfxsize=8cm\epsfbox{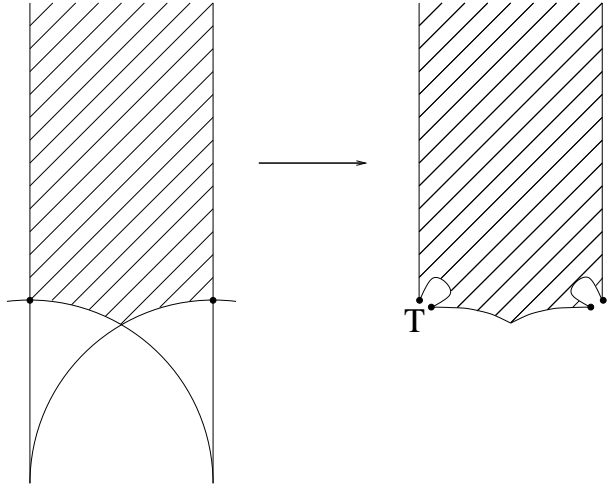}}
  \caption{The moduli space for constant $y$ for $X$.}
  \label{fig:ug}
\end{figure}
\fi

In the case of $X$ this correspondence fails. We show
schematically\footnote{We do not claim that this picture represents
the geometry very accurately. The only fact we really need is that
there is infinite monodromy at a point in the moduli space which is at
finite distance.} what happens as $y$ is switched on in figure \ref{fig:ug}.
Near the edge of the moduli space labelled by $L$ we need to continue
into a new fundamental region. This region must be distinct from that
obtained by $\tau\to\tau+1$ (otherwise there would be no monodromy
around the dots labelling the discriminant). Clearly this is
impossible without moving off onto a infinitely-sheeted cover of the
upper-half plane.

Note that one might want to extend the idea of duality by following
the route of building an infinite-sheet covering. Indeed this was the idea
behind such duality groups as those proposed in \cite{CDFvP:} for
example. One is certainly free do this mathematically and such a model
might have many uses. What is unclear however is the physical
significance of the huge Teichm\"uller space one builds by this
process. It is certainly a considerable departure for the usual
meaning of T-duality or U-duality. If we want to have a meaningful
Teichm\"uller space then we should insist that we have a unique idea
of an object at large radius and weak string coupling. This forces us
to cover our moduli space with only one copy of a plane. We will demand
this for our notion of duality.

The essential difference between this section where T-duality is
broken and the previous section where T-duality was exact is as
follows. In the unbroken case we introduced new copies of the
fundamental region close to the horizontal axis ``$\Img(\tau)=0$'' when
we switched on the coupling in figure~\ref{fig:G2a}. In this section
we had to attempt to add new regions near the point $\tau=i$. The
metric diverges near $\Img(\tau)=0$
allowing room for new regions. New regions cannot be fitted in near
$\tau=i$ where the metric is finite.

We have also seen explicitly how the argument for T-duality using
the enhanced gauge symmetry fails in this case.  The Weyl symmetry of
$\SU(2)$ was used to relate the deformations increasing and decreasing
$R$ from its value at the point of enhanced symmetry.  In fact, the
symmetry is never restored, and near the origin the monodromies about
the two singularities that do appear render any attempt to
consistently define the Weyl-noninvariant coordinate ($a$ in
\cite{SW:I}) futile.


\section{Discussion} \label{s:conc}

\subsection{Holonomy Arguments}  \label{ss:hol}

One my try to forge a general setting for T-duality (and S-duality
and U-duality) as follows. One wants to begin with a general smooth
Teichm\"uller space, $\cT$, which parameterizes familiar concepts such as 
length, coupling constants, etc. One then wants the moduli space to
look like $\cM=\cT/G$ for some discrete duality group $G$. In
particular this means that any orbifold point in $\cM$ is a {\em
global\/} orbifold point. That is, there is a global cover which
removes this singularity.

Orbifold points in a moduli space need not be global. We would like to
consider the appearance of local orbifold points as an obstruction to
duality groups. Of course one may want to declare in such a case that
the covering $\cT$ itself has the orbifold point. One is free to do
this and one ends up with a weaker notion of duality groups. In a
typical case however this weak duality group will probably be trivial
and so this is not a particularly useful notion. For duality we will
insist that {\em all\/} orbifold points are global and $\cT$ is
smooth.

Thanks to the Berger-Simons theorem (see for example chapter 10 of
\cite{Besse:E}) knowing the holonomy of a manifold can lead a
considerable knowledge of its structure. If the holonomy (on the
Levi-Civita connection) and its representation on the tangent bundle
is of a certain type then the manifold must be a symmetric space. We
will call such a holonomy ``rigid''. If the holonomy is ``non-rigid''
then the manifold is of type general Riemannian, complex K\"ahler,
hyperk\"ahler, etc.  By using the homogeneous structure of the
symmetric spaces one may argue the following by tracing geodesics.%
\footnote{We thank R.~Bryant for discussions on this.}
\begin{prop}
If a geodesically complete orbifold has a rigid holonomy then it is 
globally a quotient of a symmetric space.
\end{prop}

Note that in lower dimensions, a sufficiently extended supersymmetry
forces a rigid holonomy group.  The proposition tells us immediately
that we should expect duality groups to appear for a large number of
supersymmetries. For example for a type IIA string on a K3 surface we
have $N=2$ supersymmetry in six dimensions giving an $R$ symmetry of
$\Sp(1)\times \Sp(1)\cong\SO(4)$ (up to discrete groups). Given the
dimension of the moduli space this has rigid holonomy \cite{CFG:II}
and so we see a Teichm\"uller space in the form of a symmetric space
and a T-duality group (or more generally a U-duality group). The same
is not true for an $N=2$ theory in 
four dimensions however. Here the holonomy of the moduli space only
dictates that it be the product of a K\"ahler manifold and a
quaternionic K\"ahler manifold. This non-rigid case should not be expected
to have a duality group. This is the case of a heterotic string on
K3$\times T^2$.

This classification breaks down a little in higher dimensions.  This
happens because symmetric spaces can sometimes have a holonomy (and
representation) which appears to be non-rigid. For example, the
heterotic string on a 2-torus alone is an $N=1$ theory in eight
dimensions which has an $R$ symmetry $\GU(1)$. This tells us only that
the moduli space is K\"ahler and so would not in itself predict a
T-duality group. Nevertheless the moduli space here is a quotient of a
symmetric space. This shows that the holonomy can only tell us when we
{\em must\/} have a duality group rather than when we cannot. It seems
reasonable however in lower dimensions where 
this ambiguity does not exist that a generic compactification leading
to non-rigid holonomy will have no T-duality. 
The computation of section \ref{s:snap} shows this is the case at
least for one example.

\subsection{Consequences}  \label{ss:oops}

We have shown that when one considers a heterotic string compactified on
K3$\times T^2$ in a particular way then the factor of the moduli space
relevant to the torus is {\em not\/} a quotient of a physically
meaningful smooth space of parameters by
some T-duality group. In particular it is a completely meaningless
statement to say that there is some kind of \RoR\ symmetry on the
circles within this torus. For a nonzero value of the string coupling
constant the torus (and therefore circle) obeys a T-duality
relationship no more than a generic \CY\ threefold. The
meaninglessness of T-duality for \CY 's was discussed in
\cite{CDGP:,AGM:sd,me:min-d}.

All one may do is to begin labelling points in the moduli space with
size values near the large radius limit and then extend this process
over the entire moduli space. This will usually lead to a minimum
``size''. No real meaning can be attached to sizes below this value.

The question that immediately raises itself is the following. {\em If
T-duality fails for the heterotic string on K3$\times T^2$ then does
it also fail for a circle itself?} One might argue that if the string on
a circle of radius $R$ is exactly the same thing as a string on a
circle of radius $1/R$ then further compactification on K3$\times S^1$
should also yield identical physics. This leads to three
possibilities:
\begin{enumerate}
\item {\em T-Duality fails for the string on a circle.} That is, \RoR\
symmetry is totally wrong when all nonperturbative corrections are
taken into account. This would seem to be unlikely. For a string on a
single circle at its self-dual radius we really see an unbroken
$\SU(2)$ gauge symmetry signaling an identification in the moduli
space. This $\SU(2)$ is broken by quantum effects when we consider
K3$\times T^2$. The holonomy arguments above also show that a lot of
supersymmetry implies T-duality. We suggest therefore that T-duality
is a symmetry for a string on a circle by itself.
\item {\em Heterotic/Type II duality is wrong.} If we really want
T-duality to be respected exactly by all strings then we need to
sacrifice the duality we used in this paper. One is free to do
this. We do not know enough about string theory to ``prove'' heterotic/type
II duality. If one really is wedded to the idea of T-duality then this
is a sacrifice one might be willing to make. 
\item {\em A string doesn't know about its compactifications.} When
one says that one understands some string theory compactified on some
space, does this understanding ever include further compactification?
Such a further compactification includes putting different boundary
conditions on fields which had lived in flat space. In particular one
may break supersymmetry. The above reasoning that \RoR\ symmetry on a
circle implies T-duality for further compactification is wrong. 
\end{enumerate}

The last possibility would appear to be the most reasonable. It
implies that \RoR\ symmetry of a circle {\em depends on the context of
the circle.} Any argument which uses T-duality for a circle or torus
embedded in some more complicated situation cannot necessarily be
considered rigorous. 

In this paper we have considered the Kachru--Vafa example of the
heterotic string compactified on a torus stuck at
``$\sigma=\tau$''. This allowed us to consider just a two-parameter
family of \CY\ threefolds --- one remaining parameter for the torus
and the string coupling. Our failure to see an $\Sl(2,\Z)$ modular
group can therefore be viewed just as much as failure of the {\em
classical $\Sl(2,\Z)$ modular group\/} as of T-duality. What one
should do is to go to the three parameter example of \cite{KV:N=2} so
that we may disentangle the $\sigma$ and $\tau$ parameters of the
2-torus. The essential computation was again done in
\cite{KKL:limit}. The coarse result of this is that the moduli
space does not respect {\em any\/} $\Sl(2,\Z)$ symmetry generically.

This is perhaps a good deal more shocking at first sight than breaking
T-duality. Are we really saying that string theory breaks the
classical modular invariance of a torus? In other words string theory
does not respect diffeomorphism invariance of the target space!

This is unavoidable if one wants to maintain heterotic/type II
duality. Actually this possibility is not as bad as it might first
appear upon reflection.

One is used to the idea from conformal field theory at weak string
coupling that quantum geometry may effect the structure of the moduli
space of complexified K\"ahler form but that the complex structure
moduli space is unchanged. Now we claim that string coupling effects
may also lead to quantum corrections to the moduli space of complex
structures. This is quite reasonable in the current context as once
string corrections have been taken into account the moduli space of
the 2-torus does not even factorize exactly into K\"ahler form and
complex structure parts. Note also that the parts of the moduli space
whose shape is most effected by the corrections are those close to
where the enhanced gauge symmetry appeared in the weak coupling
limit. These must involve circles in the torus approaching the string
scale. Thus it is still distances at the string scale in some sense
which are most affected by quantum effects.

Again we should emphasize is that we are not really saying that a
string on a certain torus is different to a string on the torus having
undergone a modular transform. Rather we are saying that once the
string coupling is not zero, the moduli space of is no longer of the
form $H/\Sl(2,\Z)$. The meaning of what is it to go outside the
fundamental region then becomes unclear.

Finally we note that all of the results in this paper have tied to
dualities of the heterotic string. One might wish the type II string
to escape such effects. This appears to be unlikely. Our holonomy
argument of section \ref{ss:hol} would seem to imply that a generic
compactification with little supersymmetry should not obey any
T-duality or U-duality laws. What is true however is that the type IIA
string has more supersymmetry to begin with. Thus we know that
T-duality for the type IIA string K3$\times T^2$ {\em is\/} exact
(see, for example, \cite{AM:Ud} for a discussion of this). What one
might question however is whether the same is true for the type II
string on $Z\times S^1$ where $Z$ is a generic \CY\ threefold.


\section*{Acknowledgements}

It is a pleasure to thank P.~Argyres, R.~Bryant, S. Kachru,
J.~Harvey, G.~Moore, D.~Morrison, E.~Sharpe and E.~Silverstein for
useful discussions.


\end{document}